\documentclass[aps,prl,showpacs,twocolumn,superscriptaddress,amsmath,amssymb]{revtex4}
\usepackage{graphicx}
\usepackage{dcolumn}
\usepackage{bm}
\usepackage{subfigure}
\usepackage{color}
\usepackage{amssymb}
\usepackage{epstopdf}

\begin{document}

\title{The role of electronic excitation in cold atom-ion chemistry}


\author{Scott T. Sullivan}
\author{Wade G. Rellergert}
\affiliation{Department of Physics and Astronomy, University of California, Los Angeles, California 90095, USA}
\author{Svetlana Kotochigova}
\affiliation{Department of Physics, Temple University, Philadelphia, Pennsylvania 19122, USA}
\author{Eric R. Hudson}
\affiliation{Department of Physics and Astronomy, University of California, Los Angeles, California 90095, USA}
 
\date{\today}

\begin{abstract}
The role of electronic excitation in charge exchange chemical reactions between ultracold Ca atoms and Ba$^+$ ions, confined in a hybrid trap, is studied. This prototypical system is energetically precluded from reacting in its ground state, allowing a particularly simple interpretation of the influence of electronic excitation. It is found that while electronic excitation of the ion can critically influence the chemical reaction rate, electronic excitation of the neutral atom is less important.  It is also experimentally demonstrated that with the correct choice of the atom-ion pair, it is possible to mitigate the unwanted effects of these chemical reactions in ultracold atom-ion environments, marking an important step towards the next generation of hybrid devices.
\end{abstract}
\maketitle

Since the inception of laser-cooling, the primary focus of atomic physics has been the development of techniques for the production and study of ultracold matter -- an endeavor that, at its core, is centered on gaining full control over matter at the quantum level. This work has been extremely successful, enabling many long-sought-after goals, such as quantum degenerate gases~\cite{Ensher:1995p6055,Jin:1999p6208}, quantum simulation~\cite{Buluta:2009p6207}, quantum information~\cite{Ladd:2010p6206}, and precision measurement of fundamental physics~\cite{Bennett:1999p6269,Rosenband:2008p6279}. To gain this control, however, most ultracold matter production techniques rely on the scattering of a large number of photons from the system under study, potentially leading to a large degree of electronic excitation. Thus, the atom or molecule being cooled can be in a somewhat peculiar state: its external motion may be characterized by a temperature close to absolute zero, but its internal electronic degree of freedom may be described by a temperature approaching infinity. While this non-thermal distribution of electronic states has some notable important consequences for ultracold atoms~\cite{Walker:1989p6320}, most of its effects, such as photochemical reactions~\cite{Gould:1988p6346,Rolston:1991p6345,Sullivan:2011p6336}, are largely ignored as they occur at rates that have relatively little effect on the system.  However, as the field now moves towards producing more complex systems at ultracold temperatures, \textit{e.g.} molecules~\cite{Shuman:2010p6364} and hybrid systems~\cite{Zipkes:2010p2428,Schmid:2010p2461}, the effect of these light-assisted processes must be reevaluated as their rates may be much larger due to, among other things, an increased density of accessible product states and longer range interactions. Thus, there is presently a need to better understand the role of electronic excitation in chemical reactions at ultracold temperatures to enable the next generation of atomic physics experiments.

Interestingly, the rapidly emerging field of hybrid atom-ion systems offers a unique opportunity to study ultracold chemical reactions~\cite{Grier:2009p2279,Zipkes:2010p2430,Rellergert2011,Hall2011}. Like the all-neutral systems of traditional atomic physics, the two trapped species can collide at short-enough range for chemical reactions to proceed; but, unlike all-neutral systems it is possible to maintain a product of the chemical reaction in the trap, since ion trap depths are large relative to the kinetic energy gained in most chemical reactions. Already, such hybrid systems have been used to measure several important ion-neutral chemical reactions \cite{Rellergert2011,Grier:2009p2279,Zipkes:2010p2430,Schmid:2010p2461}, culminating in the recent observation of molecular ion reaction products \cite{Hall2011}.  Here, we utilize a hybrid atom-ion MOTION trap system~\cite{Rellergert2011,Hall2011} to study reactions between ultracold $^{40}$Ca atoms and $^{138}$Ba$^+$ ions. Unlike all previous studies,  the only energetically accessible chemical reaction pathways for this system require the atom, the ion, or both to be electronically excited. As this reaction takes place between a closed-shell atom and open-shell ion, the results of this system are simpler to interpret and serve as a general prototype for drawing conclusions about more complicated systems.  Further, by utilizing the richer electronic structure of $^{40}$Ca, as compared to neutral atoms used in other hybrid systems, we are able to vary the population in the neutral atom electronic states over a wider range than in previous experiments and present the first detailed experimental investigation of the importance of neutral atom excitation in ion-neutral chemistry.

In what follows, we briefly describe our experimental system, discuss how optical coherence and trap dynamics can critically affect the observed atom-ion chemical reaction rate when electronic excitation is involved, and present experimental results detailing the observed chemical reaction pathways and the average radiative association branching ratio for formation of CaBa$^{+}$ molecular ions. We also present relativistic \textit{ab initio} molecular structure calculations, which further illuminate the chemical reaction pathways as well as provide guidance for future experiments. We conclude with a demonstration that it is possible to avoid the large chemical reaction rates seen in other hybrid atom-ion systems, with reasonable experimental parameters, marking a crucial step towards the next generation of hybrid atom-ion devices.

\begin{figure*}
\resizebox{2\columnwidth}{!}{
    \includegraphics{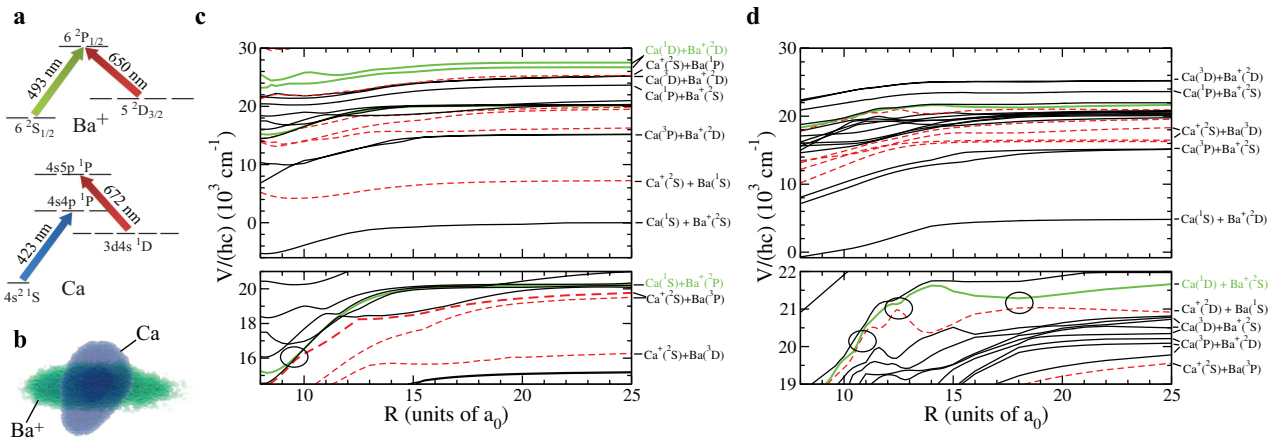}
}
\caption{a) Ba$^+$ and Ca relevant energies levels and laser cooling schemes. b) Fluorescent images of Ca MOT and Ba$^+$ ion cloud. c) and d) Molecular potentials of the $\Omega$=$\frac{1}{2},\frac{3}{2}$ excited states of the BaCa$^+$ molecule, respectively, as a function of interatomic separation $R$. The bottom panels show higher detail for potentials near $20\times 10^3$ cm$^{-1}$ above the Ca$^+$($^2$S)+Ba($^1$D) dissociation limit. Entrance and exit channels, as defined in the text, dissociate to green and red potentials, respectively.}
\label{ReactionChannels}
\end{figure*}

The apparatus used in this work is similar to that described in Ref.~\cite{Rellergert2011}. Briefly, $^{138}$Ba$^+$ ions are loaded into a linear radio-frequency quadrupole trap via laser ablation of a BaCl$_2$ target~\cite{Chen2011} and cooled with colinear laser beams.  In the same region, ultracold Ca atoms are produced and trapped in a magneto-optical trap (MOT). (See Fig.~\ref{ReactionChannels}(a) for the respective ion and atom laser cooling schemes.)  Two nearly-orthogonal cameras allow for a 3-dimensional reconstruction of both the Ca MOT and the laser-cooled ion cloud, and are used to quantify the degree of overlap between the Ca atoms and  $^{138}$Ba$^+$ ions, as shown in Fig.~\ref{ReactionChannels}(b). Using the same approach as Ref.~\cite{Rellergert2011}, the laser-induced 493~nm fluorescence monitors the $^{138}$Ba$^+$ ion population, which is observed to decay in the presence of the Ca MOT due to charge exchange chemical reactions.  Special care must be taken to account for the time evolution of the degree of overlap, which requires a non-analytical numerical fitting technique, described in Appendix A, to account for the deviation from simple exponential decay.

In this experiment, the atom-ion collision energies are primarily set by the ion micromotion since the temperature of the Ca MOT (T~$\simeq~4$~mK) and the Doppler-cooling temperature limit of the Ba$^{+}$ ions  (T~$\simeq~0.5$~mK) are both lower than the typical, position dependent, energy of micromotion in the ion cloud ($\frac{E_{\mu}}{k_{B}}\sim$10~mK-10~K). At these low collision energies, chemical reaction dynamics are primarily determined by the relative energies of the entrance and exit reaction channels, shown in black and red, respectively, in Fig.~\ref{ReactionChannels}(c,d). Because the ionization potential of Ca (6.1~eV) is significantly larger than that of Ba (5.2~eV), charge exchange reactions between the Ca 4$^1$S- and Ba$^+$ 6$^2$S- or 5$^2$D- electronic states are energetically precluded. As a result, chemical reaction between Ba$^+$ and Ca can only occur if the atom, the ion, or both are electronically excited. Of the ten energetically allowed entrance channels populated in this experiment, the six involving the short-lived Ca P-states do not contribute to the observed reaction rate. Because the Ca 4$^1$S-, 4$^1$P- and 5$^1$P-states have very different atomic polarizabilties~\cite{Mohammadou} (163.0~a.u., 55.3~a.u., and $>$1000~a.u., respectively), during the collision the strong monopole field of the ion alters the atomic transition frequency such that the Ca MOT lasers are shifted out of resonance for atom-ion separations of many hundred Bohr radii, $a_o$. Thus, though a collision leading to chemical reaction may begin with a Ca atom in a P-state, by the time the atom and ion are close enough together to react ($<10~a_o$), the Ca atom has radiatively decayed from the P-state. We note similar effects are discussed in Ref.~\cite{Band:1992p3861,Wallace1995,Grier:2009p2279}, were observed in Ref.~\cite{Rellergert2011}, and are confirmed by the data presented here. This effect appears generic and it is likely that short-lived excitations of ultracold atoms can be ignored in future experiments involving reactions of ultracold atoms and ions. (This effect is particularly beneficial to techniques for sympathetic cooling of molecular ions by ultracold atoms~\cite{Hudson:2009p1235}, as it significantly relaxes the experimental constraints.) Interestingly, a similar argument cannot be made for short-lived, excited states of ions. Because excitation of the ion only changes the long-range dispersion coefficient, e.g. C$_{6}$, the ion cooling lasers are not shifted out of resonance until relatively small atom-ion separations. Thus, while these chemical reactions are somewhat suppressed, an effect that must be accounted for when evaluating rate constants involving excited states susceptible to spontaneous decay, it is nonetheless possible for them to occur between atoms and excited state ions.  

With these considerations in hand, it is clear that the important reaction pathways proceed from the energetically allowed entrance channels involving 4$^1$S- or 3$^1$D-state Ca atoms and 6$^2$S-, 6$^2$P-, or 5$^2$D-state Ba$^+$ ions, highlighted in green in Fig.~\ref{ReactionChannels}(c-d). While the total charge exchange rate constant for each of the four distinct pathways depends sensitively on the details of the CaBa$^+$ molecular structure, it can nevertheless be measured in a straightforward way by simply varying the excited state populations of the atoms and ions while recording the total system reaction rate. Experimentally, this variation is accomplished by changing the intensities of the atom and ion cooling and repumping lasers. From the known experimental parameters it is then possible to calculate the relative population of each atom and ion electronic state, and extract the contribution of each reaction channel to the total measured reaction rate.

 \begin{figure}
\resizebox{1\columnwidth}{!}{
    \includegraphics{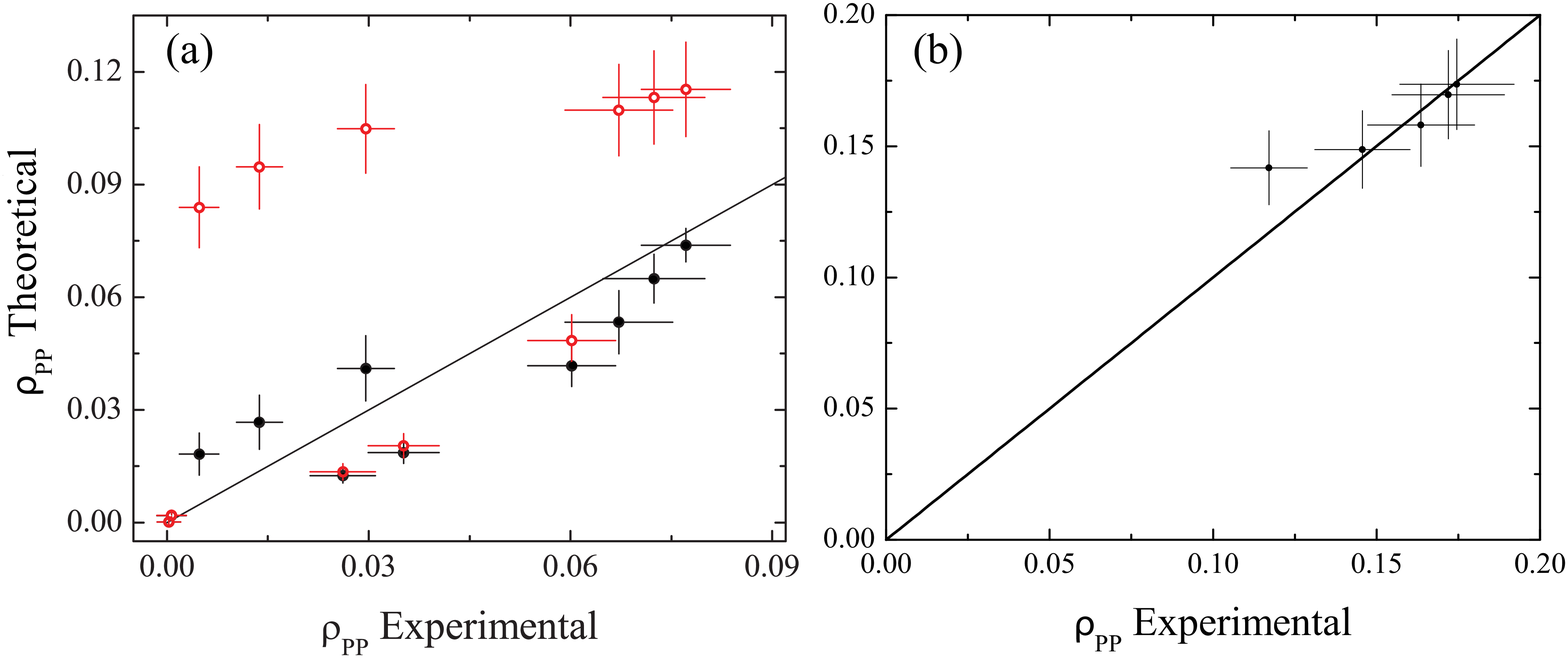}
}
\caption{Panel (a): experimentally determined excitation fraction in Ba$^{+}$ versus theoretical predictions of a density matrix ($\bullet$) and rate equation (\textcolor{red}{$\circ$})  treatment. Panel (b): experimentally determined excitation fraction in Ca versus predictions of a 4-level rate equation model.  In both panels the solid line has a unit slope and represents perfect agreement between experiment and theory.}
\label{OC}
\end{figure}

The calculation of the Ba$^+$ electronic state populations is significantly complicated by optical coherence effects and the quadrupole magnetic field of the MOTION trap, whose direction and strength vary in a non-trivial way. Therefore, we determine the Ba$^+$ electronic state populations by solving the steady-state quantum Liouville equation for an eight-level optical Bloch Hamiltonian at each position in the ion cloud and spatially averaging the result, as described in Appendix B. Similarly, a first principles calculation of the electronic state populations of atoms in a MOT environment is a well-known open problem and various phenomenological models have been proposed~\cite{Shah2007, Javanainen1993}. Thus, to verify the calculation of Ba$^+$ populations and determine the appropriate model for Ca populations, we experimentally measure the fraction of population in both the Ba$^+$ 6$^{2}$P$_{1/2}$ state, $\rho^{Ba}_{pp}$, and the Ca 4$^1$P-state, $\rho^{Ca}_{pp}$, by comparing the observed fluorescence rates of each species to the total number of scatters as measured by a channeltron ion detector and absorption imaging, respectively. The results of the Ba$^+$ population measurement are summarized in Fig.~\ref{OC}(a), where the predictions of a simple rate equation model are included for comparison. The poor agreement of the $\rho^{Ba}_{pp}$ predicted by the rate equation model highlights the need for the density matrix treatment, as its use can lead to significant underestimate of the charge exchange reaction rate constant in three-level systems.  The results of the Ca population measurement are shown in Fig.~\ref{OC}(b) along with the prediction of a simple four-level rate equation model. The good agreement with the model is likely a consequence of the lack of hyperfine structure in the $^{40}$Ca system; however, in the absence of a first principles model, it could simply be fortuitous. Using this result, we estimate the fraction of Ca in the 3$^1$D-state, $\rho^{Ca}_{dd}$, by finding equilibrium between the 4$^1$P$\rightarrow$3$^1$D spontaneous decay rate and the combination of the ballistic exit rate from the MOT and laser repumping rate~\cite{Sullivan:2011p6336}.

 \begin{figure}
\resizebox{1\columnwidth}{!}{
    \includegraphics{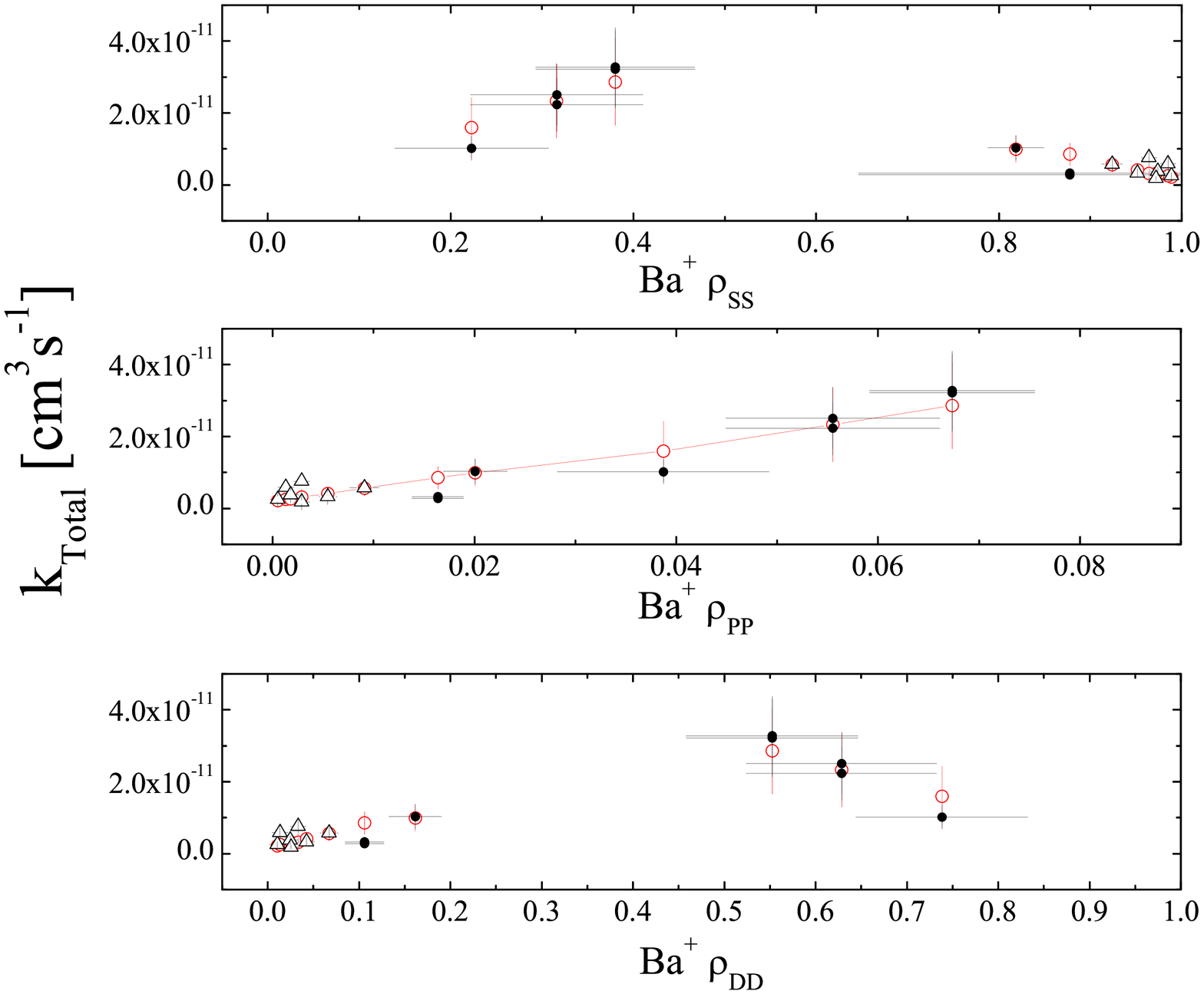}
}
\caption{Measured rate constant as a function of Ba$^+$ population fractions.  Experimental result in black.  Best fit to the four entrance channel model shown as open red circles (\textcolor{red}{$\circ$}).  
}
\label{RatesVsIon}
\end{figure}

 \begin{figure}
\resizebox{1\columnwidth}{!}{
    \includegraphics{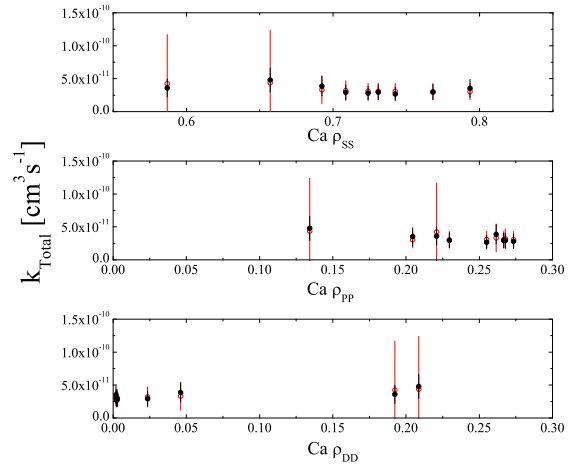}
}
\caption{Measured rate constant as a function of Ca population fractions.  Experimental result in black.  Best fit to the four entrance channel model shown as open red circles(\textcolor{red}{$\circ$}). 
}
\label{RatesVsMOT}
\end{figure}

With these tools in place, we measure the reaction rate constant as a function of the Ba$^+$ and Ca excited state fraction, as shown in Figs.~\ref{RatesVsIon} and \ref{RatesVsMOT}, respectively.   Care must be taken in interpreting the measurements presented in these graphs due to the correlations among the fractional populations as the laser intensities are varied. As a function of Ba$^+$ excitation, the most obvious feature is a linear trend of increasing reactivity with increasing $\rho^{Ba}_{pp}$.  In addition, the reaction rate constant diminishes at both extremes of the Ba$^+$ S-state,  $\rho^{Ba}_{ss}$, and D-state, $\rho^{Ba}_{dd}$, population fraction. Taken together, these features indicate that the primary reaction channel involves the Ba$^+$ ion P-state. Also shown in Fig.~\ref{RatesVsIon}, as open triangles, are several measurements at low excitation fractions ($<1\%$). While these points appear to qualitatively agree with the rest of the data, we report data analysis both with and without inclusion of these points in order to avoid possible systematic inaccuracies due to higher secular energies.

The measurement of the rate constant over the experimentally accessible Ca excitation range, Fig.~\ref{RatesVsMOT}, shows no detectable variation for fixed Ba$^+$ trapping parameters. With the aforementioned assumption that Ca atoms in the 4$^1$P-state radiatively decay to the 4$^1$S-state by the time they are close enough to an ion to chemically react and with negligible 5$^1$P-state population for our operating parameters, this data indicates that the Ca S-state and D-state exhibit similar reaction rates with Ba$^+$ ions. However, since the four-level rate equation model has only been verified to accurately predict the fraction of Ca atoms in the P-state (experimental verification of the D-state populations in not presently possible in our apparatus), the possibility remains for systematic errors associated with an inaccurate calculation of the Ca D-state population. 

To extract channel-specific reaction rate constants from these data, we quantify the total reaction rate constant in terms of the excited state fractions of both Ba$^+$ and Ca as:
$k = \rho^{Ba}_{pp} \rho^{*Ca}_{ss} k_{ps}+ \rho^{Ba}_{ss} \rho^{Ca}_{dd} k_{sd}+ \rho^{Ba}_{dd} \rho^{Ca}_{dd} k_{dd}+ \rho^{Ba}_{pp} \rho^{Ca}_{dd} k_{pd}.$
By accounting for the conservation of probability ($\rho_{ss} = 1 - \rho_{pp} - \rho_{dd}$) and the fact that Ca atoms in the P-state radiatively decay to the S-state before the atom and ion are close enough to react ($\rho^{*Ca}_{ss} = \rho^{Ca}_{ss} + \rho^{Ca}_{pp}$), the totality of the data in Fig.~\ref{RatesVsIon} and Fig.~\ref{RatesVsMOT} was subject to a multi-parameter fit along the manifold $\left\{\rho^{Ba}_{pp},\rho^{Ba}_{dd},\rho^{Ca}_{dd}\right\}$ to determine the contribution of each reaction pathway to the total rate constant. The reaction rate constant of the primary entrance channel, Ba$^+$(6$^2$P$_{1/2}$) + Ca(4$^1$S), was measured to be $k_{ps}= 4.2(1.9)\times10^{-10}$~cm$^{3}$~Hz.  The constraints for the other entrance channels vary depending on whether the low excitation fraction Ba$^+$(6$^2$P$_{1/2}$) data is included in the fits.  If the small excitation population points are included in the analysis, the rate constant for the only channel out of the Ba$^{+}$ ground state, Ba$^+$(6$^2$S$_{1/2}$) + Ca(3$^1$D), is poorly constrained with a best fit of $k_{sd} \approx 2.0\times10^{-10}$~cm$^{3}$ Hz.  If the points are not included, the best fit for $k_{sd}$ becomes consistent with zero within a large standard error of $ \approx 3.0\times10^{-10}$~cm$^{3}$ Hz.   Because the remaining two entrance channels, Ba$^+$(6$^2$P$_{1/2}$) + Ca(3$^1$D) and Ba$^+$(5$^2$D$_{3/2}$) + Ca(2$^1$D), require excitation of both the atom and ion our experiment is relatively insensitive to them and we can only set upper limits on them of $k_{pd} \leq2\times10^{-9}$~cm$^{3}$~Hz and $k_{dd} \leq7\times10^{-10}$~cm$^{3}$~Hz. This fitting method yields a surface in dimensions of the various excitation fractions which was traversed in a non-trivial manner under variation of laser intensities, and thus it is not possible to plot a line of best fit.  Therefore to show the quality of agreement between the best fit model and the results, we plot a data point generated from the model corresponding to each experimental value (open circle, red), with error bars including the full uncertainty associated with the upper  bounds of the doubly excited channel. The relatively uncertain agreement for the measurements taken at higher Ca D-state population is indicative of either a failure in our estimation of the D-state fraction or that the upper bounds are too conservative.

To illuminate the charge exchange mechanisms responsible for the individual reaction pathways, we performed {\it ab initio} calculations of the $\Omega$=1/2, 3/2  potentials using a relativistic multi-reference restricted active space  configuration-interaction method \cite{Kotochigova}. Spin-orbit effects are large in CaBa$^+$ and a relativistic calculation is required as described in Appendix~C. The resulting Born-Oppenheimer potentials for the ground and first-excited state of CaBa$^+$ are shown in Fig.~\ref{ReactionChannels}(c-d), and assigned by their atomic dissociation limit.  
The bottom panels of Fig.~\ref{ReactionChannels}(c-d) show a higher resolution view of the closely spaced potentials above $20\times 10^3$ cm$^{-1}$.  Both panels reveal that there are strong interactions and avoided crossings between neighboring excited potentials.  Potentials dissociating to energies $>27\times 10^3$ cm$^{-1}$ are not shown, as the exceedingly high level density makes it impossible to resolve the avoided crossings with our calculation.

In principle, a coupled-channels calculation that includes the mixings between these potentials due to radial non-adiabatic coupling and spontaneous emission, like that performed in Ref.~\cite{Rellergert2011}, can be used to calculate the non-radiative, radiative, and radiative association charge exchange reaction rate constants for each entrance channel; however, due to the high level density in the excited CaBa$^+$ molecule, this calculation is extremely technically demanding and falls outside the scope of this work. Nonetheless, the basic mechanisms of each charge exchange pathway can be inferred from the structure of its entrance channel molecular potential. For example, the experimentally determined primary entrance channel, Ca(4$^1$S) + Ba$^+$(6$^2$P$_{1/2}$), exhibits strong interactions with several potentials as well as a narrowly-avoided crossing with a molecular potential dissociating to the Ca$^+$(4$^2$S) + Ba(6$^3$P) exit channel.  The second experimentally addressed entrance channel, Ca(3$^{1}$D) + Ba$^+$(6$^2$S$_{1/2}$), has several strong avoided crossings with the Ca$^+$(3$^{2}$D) + Ba(6$^1$S) exit channel.  These qualities imply that non-radiative charge exchange is most likely responsible for the observed reaction rate.  By a similar analysis, it is expected that the Ca(3$^1$D) + Ba$^+$(5$^2$D) entrance channel would be relatively inefficient for charge exchange reactions since low energy scattering events do not experience an avoid crossing nor do they couple well to states accessible via spontaneous emission.  The remaining Ca(3$^1$D) + Ba$^+$(6$^2$P$_{1/2}$) entrance channel resides in a host of densely packed levels, which both prevents our \textit{ab initio} calculation from providing reliable results and suggests that non-radiative and radiative chemical reactions could proceed very quickly from these states, a result that is consistent with the limit set for this channel by our measurement. 

Despite the fact that the rate constants measured here are large by atomic physics standards, roughly a third of the Langevin rate, with proper engineering of the atomic and ionic excited state populations their effects can be mitigated. By using large repumping laser powers to maintain a negligible Ca D-state fraction and using the minimum ion laser cooling power necessary to achieve Doppler-limited cooling, we have observed Ba$^+$ ion lifetimes in excess of several minutes in the presence of the MOT.
Finally, using the method described in Ref.~\cite{Rellergert2011}, we measure the molecular ion product branching ratio to be $\sim 0.14$.  Future measurements, using time-of-flight detection as in Ref.\cite{schowalter2012revsci}, are planned to further constrain the branching ratio.

In summary, we have experimentally and theoretically studied chemical reactions in a hybrid atom-ion system, where only electronically excited states are permitted to react. The observed reaction dynamics are qualitatively different than what is observed in systems where ground-state reactions dominate~\cite{Rellergert2011}. The measured reaction rate constant is found to be an appreciable fraction of the Langevin rate and most-likely due to non-radiative charge transfer occurring at narrowly-avoided crossings in at least two of the entrance channels. We have also demonstrated that while special care must be taken to account for \textit{e.g.} optical coherence and trap dynamics when understanding the degree of electronic excitation, it is possible to engineer experiments where the cold reactive species can co-exist for several minutes. From these results it is clear that the combination of ultracold atoms with high ionization potential and ultracold ions with low electron affinities, along with careful control of electronic excitation should enable a new generation of hybrid-atom ion devices, capable of the long coherence times needed for proposed hybrid atom-ion systems~\cite{Schneider:2010p2516, Daley:2004p5939, Schuster:2011p5946, Cote:2000p5997, Smith:2005p6049, Hudson:2009p1235}.

This work was supported by NSF grant No. PHY-1005453, ARO grant No. W911NF-10-1-0505 and AFOSR grant No. FA 9550-11-1-0243. 

\bibliography{Ba_CEX_Bib}
\appendix
\section{Appendix A: Time-varying overlap factor } \label{app:TimeVaryingOverlapAppendix}
As the Ca MOT exhibits non-uniform particle densities on length scales comparable to the size of laser-cooled ion cloud, it is necessary to carefully consider the effective density of reactants when evaluating the reaction rate constant. Further, because the chemical reaction leads to loss of ions, the ion cloud size shrinks during the reaction process, resulting in a time-variation of this effective density. This effect can cause various systematic shifts in the measurement of the rate constant if not taken into account. For example, if the ion and atom clouds are nonconcentric, as the ion cloud shrinks the observed reaction rate slows as a result of the reduced reactant density.
Thus, to properly measure the reaction rate constant, we explicitly include the time evolution of the effective reactant density via a term that quantifies the spatial overlap of the reactants, defined as $\phi(t) = \int \hat{\rho}_{Ca} (\vec{r}) \bar{\rho}_{I}(\vec{r}) d\vec{r}$, where $\hat{\rho}$ denotes a peak normalized density and $\bar{\rho}$ denotes an integral normalized density. With this definition, the evolution of the number of Ba$^+$, $N$, ions is given as:
\begin{align}
\frac{dN}{dt} = - k \rho \phi(t) N
\end{align}
where we take the product of the reaction rate constant, $k$, to the peak neutral density, $\rho$, to be the unbiased reaction rate constant $\Gamma_{\rm{CEX}}$.

\begin{figure}
\resizebox{0.8\columnwidth}{!}{
    \includegraphics{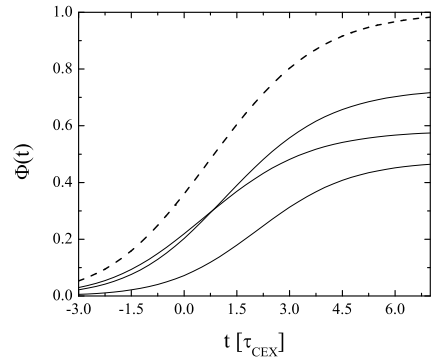}
}
\caption{The dashed line shows the numerical calculation of the overlap factor as a function of charge exchange lifetimes expired in the case of perfect alignment of relative MOT to ion cloud position, with typical experimental waists and aspect ratio.  The other three lines are randomly selected numerical calculations from actual experimental data to show how imperfect alignment changes the curve.}
\label{OverlapVersusTimeExample}
\end{figure}

In principle, $\phi(t)$ can be found numerically from measurements of the cloud widths and relative positions via dual camera fluorescence imaging at various times. 
However, due to disparate levels of brightness, the MOT and the ions cannot be simultaneously imaged with our current image filtering systems. Therefore, in practice, we measure the relative position of the two clouds and their initial density profiles. With the assumption that the Ca MOT has a roughly spherical gaussian density distribution, while the ion cloud has an ellipsoidal (cigar-shaped) gaussian density distribution, we can then uniquely determine numerically the time evolution of the overlap factor  (in units of ${\Gamma_{\rm{CEX}}^{-1}}$), as in Fig.~\ref{OverlapVersusTimeExample}.  Thus, for the determination of $\Gamma_{\rm{CEX}}$ for each decay measurement, a numerical function describing the overlap factor is uniquely determined from the images so that the only physical fitting parameter is $\Gamma_{\rm{CEX}}$.  That is, each experimental decay curve is effectively fit to $\frac{dN}{dt} = - \Gamma_{\rm{CEX}} \phi(\Gamma_{\rm{CEX}} t) N$ with an uniquely determined $\phi$.  In order to use numerical fitting algorithms, we take the solution to the differential equation to be $N(t)~\sim \rm{exp}(-\Gamma_{\rm{CEX}} \phi(\Gamma_{\rm{CEX}} t))$.  This amounts to an assumption that the overlap factor is slowly varying on the timescale of the reaction and for the data reported here leads to systematic error $\leq30\%$, which is well within the reported experimental error.

\section{Appendix B: Calculation of Ba$^+$ fractional electronic state populations}
Because the trapped Ba$^+$ ions are laser cooled via a closed three-level electronic system, see Fig.~\ref{ReactionChannels}(a), it is necessary to account for optical coherence effects to accurately calculate the excited state population fractions realized in the experiment. This is accomplished by solving for the steady-state solution to the quantum Liouville equation for an eight-level optical Bloch Hamiltonian. This calculation is significantly complicated for the MOTION trap due to the presence of a large quadrupole magnetic field  necessary for the operation of the Ca MOT as its direction and strength vary in a non-trivial way. We briefly outline this calculation in this appendix.

Working under the electric dipole approximation in the interaction picture, the eight bare $|\{LS\},J,M_J\rangle$ states have one of three energies, $E_{S} = 0,E_{P} = \delta,E_{D} = \Delta$, where $\delta$ and $\Delta$ are the one- and two-photon detunings, respectively. To simplify the geometry of the problem, we define the quantization axis to be parallel to the direction of the linearly polarized laser used to drive the Ba$^+$ $5^2$P$_{1/2}\leftarrow5^2$S cooling transition, which is approximately vertical in the laboratory frame. The laser used for population repumping on the $5^2$P$_{1/2}\leftarrow5^2$D$_{3/2}$ transition is collinear with this cooling laser and is linearly polarized at an angle $\theta$ with respect to the z-axis. As a result, the Rabi frequencies for transitions between the different $\left|\{LS\}J,M_J\right>$ states of the Ba$^+$ ion are given as:
\begin{align}
&\Omega_{L',L}(J',M_J',J,M_J,E_{1,q}) = \\ \nonumber
&(-1)^{J' - M_J'}  \left(
  \begin{array}{ccc}
    J' & 1 & J \\ \nonumber
    M_J' & q & M_J
  \end{array}
\right) \langle L',J'||\vec{d}|| L,J \rangle \mathcal{E}_{1,q}/\hbar
\end{align}
where the index $L$ = $^2$S, $^2$P$_{1/2}$, $^2$D$_{3/2}$ and $ \mathcal{E}_{1,0} =  \mathcal{E}_z$ and $ \mathcal{E}_{1,\pm1} = \pm\frac{1}{\sqrt{2}}( \mathcal{E}_x \mp \imath  \mathcal{E}_y)$ are the spherical irreducible tensor representation of the electric fields of the laser resonant with the $L'_{J'}\leftarrow L_J$ transition -- with the defined geometry, $ \mathcal{E}_x= \mathcal{E}_y=0, \mathcal{E}_z = \sqrt{\frac{2 I_{PS}}{\epsilon_o c}}$ for the cooling laser and is $ \mathcal{E}_x =  \sqrt{\frac{2 I_{PD}}{\epsilon_o c}} \sin\theta, \mathcal{E}_y = 0,  \mathcal{E}_z = \sqrt{\frac{2 I_{PD}}{\epsilon_o c}} \cos\theta$ for the repumping laser, where $I_{PL}$ is the intensity of the laser driving the transition. The angle $\theta$ is experimentally chosen to mitigate the effects of coherent population trapping and, in our experiment, is typically $\theta \approx \pi/3$.

Likewise, the effect of the spatially varying magnetic field is calculated by including in the Hamiltonian the diagonal and off-diagonal Zeeman matrix elements:
\begin{align}
& \left<J',M_J'\right|-\vec{\mu}\cdot \vec{B}\left|J,M_J\right> = \\ \nonumber
&-\frac{e\hbar}{2 m_e} g(J) \left(\frac{J_+B_{1,-1}}{\sqrt{2}}-\frac{J_-B_{1,1}}{\sqrt{2}} + J_oB_{1,0}\right)
\end{align}
where $B_{1,0} = B_z$, $B_{1,\pm1} = \mp\frac{1}{\sqrt{2}}(B_x \pm \imath By)$, and $g(J) = (3J(J+1)+S(S+1)-L(L+1))/(2J(J+1))$. Over the spatial extent of the Ba$^+$ ion cloud the magnetic field is well approximated as $\vec{B} = \frac{\delta B}{\delta z}(\frac{x}{2}\hat{x} + \frac{y}{2}\hat{y} +  z\hat{z})$, with $\frac{\delta B}{\delta z} = 0.7$~T/m for our experimental parameters.  Here we ignore off-diagonal Zeeman couplings between different electronic states, e.g. matrix elements between P$_{3/2}$ and P$_{1/2}$ states.

Finally, the effects of spontaneous emission and the resulting decoherence are included via the Lindblad super-operator and result in a relaxation matrix, $\Gamma_r$, whose diagonal elements account for the decay and growth of electronic state populations at the respective spontaneous emission rate and whose off-diagonal elements account for the concomitant decay in coherence at half of the spontaneous emission rate.

With all of these terms taken together, the optical Bloch Hamiltonian, 
\begin{equation}
\imath\hbar \frac{d\rho}{dt} = \left[H(\vec{x},I_{PS},I_{PD},\theta, \delta,\Delta),\rho\right] + \imath\hbar \Gamma_{\rm{r}}, 
\end{equation}
is then solved numerically for given experimental conditions, i.e. a given set of $\{I_{PS},I_{PD},\theta, \delta,\Delta\}$, to find the steady state Ba$^+$ populations at each location in the ion cloud. A spatial average is then performed over these values to give the average electronic state populations of the Ba$^+$ ions for the experimental conditions. For the conditions studied in this work, this technique gives excellent agreement with measured values of Ba$^+$ electronic populations, as evidence in Fig.~\ref{OC}(a). 
 
\section{Appendix C: ab initio calculation}
This RMR-RAS-CI method is based on partitioning the occupied and unoccupied molecular orbitals into subsets, which define the construction of the molecular wave function. These symmetry subsets are: core orbitals, which do not participate in the CI procedure; valence orbitals, which are occupied and can have single, double, and triple excitations; and virtual or unoccupied orbitals, which have double excitations. Though the inactive core orbitals do not contribute to correlation effects, they create, a so-called, ``self-consistent-field sea" in which the other electrons move. The valence and virtual orbitals are correlated in the CI expansion. Since the dimension of the CI molecular wave function increases dramatically with the size of the active orbital space, it is necessary to apply restrictive measures.  

A non-orthogonal basis set is constructed from numerical Dirac-Fock atomic orbitals as well as relativistic Sturmian functions. A symmetric re-expansion of atomic orbitals from one atomic center to another simplifies the calculation of many-center integrals. At large interatomic separations the molecular wave function has a pure atomic form that appropriately describes the molecular dissociation limit. The $1s^2 2s^2 2p^6 ...4d^{10}$ closed shells of barium and $1s^2 2s^2 2p^6$ closed shells of calcium atom are included in the core. An R-dependent  all-electron core potential is calculated exactly and included in the Hamiltonian. The core-valence basis set is constructed from Dirac-Fock core $5s$ and $5p$ orbitals for $Ba$ and $3s$ and $3p$ orbitals for Ca.  Valence Dirac-Fock orbitals are $6s$, $5d$, and $6p$ for Ba and $4s$, $3d$, and $4p$ for Ca, while Sturm virtual $7s$, $7p$, $6d$, $8s$ $8p$ and $7d$ orbitals for Ba and $5s$, $5p$, $4d$, $6s$, $6p$, and $5d$ orbitals for Ca are also included. 

\end{document}